\documentclass{article}
\usepackage[utf8]{inputenc}
\usepackage{authblk}
\usepackage{indentfirst}
\usepackage{hyperref}
\usepackage{longtable}
\usepackage{setspace}
\usepackage[margin=1in]{geometry}
\usepackage{graphicx}
\graphicspath{ {./figures/} }
\usepackage{subcaption}
\usepackage{amsmath}
\usepackage{lineno}
\usepackage{multirow}
\usepackage{color,soul}
\usepackage[version=3]{mhchem}

\usepackage[biblabel]{cite}

\title{Leveraging transfer learning for accurate estimation of ionic migration barriers in solids}

\author[1]{Reshma Devi}
\author[2,*]{Keith T. Butler}
\author[1,*]{Gopalakrishnan Sai Gautam}

\affil[1]{Department of Materials Engineering, Indian Institute of Science, Bengaluru, 560012, India}
\affil[2]{Department of Chemistry, University College London, London WC1E 6BT, United Kingdom}
\affil[*]{Email: \href{mailto:k.t.butler@ucl.ac.uk}{k.t.butler@ucl.ac.uk}; \href{mailto:saigautamg@iisc.ac.in}{saigautamg@iisc.ac.in}}

\date{}

\begin{document}

\maketitle

\begin{abstract}
Ionic mobility, which determines the rate performance of several applications, such as batteries, fuel cells, and electrochemical sensors, is exponentially dependent on the migration barrier ($E_m$) for ionic motion within solids, a quantity that is difficult to measure experimentally or estimate computationally. Previous approaches to identify materials with high ionic mobility have often relied on imprecise descriptors or rules-of-thumb given the lack of generalizable models to predict $E_m$ swiftly and accurately. Here, we present a graph neural network based architecture that leverages principles of transfer learning to efficiently and accurately predict $E_m$ across a diverse set of materials. We use a model (labeled MPT) that has been simultaneously pre-trained on seven distinct bulk properties, introduce modifications into the MPT model to build inductive bias on different migration pathways in a structure, and subsequently fine-tune (FT) on a manually-curated literature-derived dataset of 619 $E_m$ data points calculated with density functional theory. Specifically, we propose four different architectural modifications of the MPT model that differ on the number of instances of the MPT model used, the structural images or bands taken as an input, and the presence/absence of attention layers. Importantly, our best-performing FT model (labeled MODEL-3) demonstrates substantial improvements in prediction accuracy compared to classical machine learning methods, graph models trained from scratch, and a universal machine learned interatomic potential, with a R$^2$ score of 0.703 and a mean absolute error of 0.261~eV on the test set. Notably, MODEL-3 is able to distinguish different migration pathways within a given structure and also demonstrates excellent ability to generalize across intercalant compositions and chemistries. As a classifier, MODEL-3 exhibits 80\% accuracy and 82.8\% precision in identifying materials that are `good' ionic conductors (i.e., structures with $E_m <$0.65~eV). Thus, our work demonstrates the effective use of FT strategies, architectural modifications necessary to learn a target property, and the adaptability of our MPT model, which can be extended to make predictions on other data-scarce material properties. Finally, the demonstrated abilities of MODEL-3 highlights its potential use in identifying materials with high ionic mobility with applications in batteries and beyond. 
\end{abstract}


\section{Introduction}
Ionic conductivity ($\sigma$) or diffusivity ($D$) or mobility ($\mu$) in a crystalline solid is an important material property that governs the rate performance of several applications, such as rechargeable batteries, fuel cells, and electrochemical sensors. In the case of secondary batteries, the core operating principle involves the reversible movement of mobile ions (such as Li$^{+}$, Na$^{+}$, K$^{+}$, Mg$^{2+}$, or Ca$^{2+}$) between electrode materials that are typically intercalation compounds and across an electronically-insulating electrolyte, which facilitates electron exchange via the external circuit.\cite{simon2014batteries} Thus, the rate performance of a given rechargeable battery is often limited by the ionic mobility of the electroactive ion in the intercalation electrode and/or in the solid electrolyte (in case of all-solid-state batteries).\cite{bachman2016inorganic,sotoudeh2024ion} Thus, significant improvements in rate performance of batteries (and other electrochemical applications) can be effected by the use of novel materials\cite{griesemer2023accelerating} that exhibit high $\mu$ of the electroactive ion, indicating that computational or experimental screening strategies\cite{maffettone2021crystallography} need to predict $\mu/\sigma/D$ swiftly and accurately for the identification of candidate materials. 

Mathematically, $\mu$ and $\sigma$ are directly proportional to one another, with $\sigma$ usually described by the Nernst-Einstein equation 
$(\sigma = \frac{q^2 x D(x)}{k_B T})$, where $q$, $x$, $k_B$, and $T$ are the charge and concentration  of the intercalant, the Boltzmann constant, and temperature, respectively. $D(x)$ relates the diffusive flux to the concentration gradient via Fick's first law,\cite{fick1855v} and can be expressed as $D(x) = D_J(x) \theta(x)$. $D_J$ is the jump diffusion coefficient that captures the rate of atomic migrations and the correlations among atomic hops, and $\theta$ is the thermodynamic factor that accounts for any non-ideal interactions between the intercalant and the diffusion carrier (typically vacancies). Thus, macroscopic diffusion of ions in a solid ($D$) is directly related to microscopic atomic hops ($D_J$) that happen within the crystal structure. In the case of ideal interactions between the intercalant and vacancies, where each atomic hop exhibits an identical frequency (or probability) of occurrence, $D$ simplifies to Equation~{\ref{eq:diffusion_coefficient}}. 
\begin{equation} \label{eq:diffusion_coefficient}
D = f.g.a^2.\nu.\exp\left(-\frac{E_m}{k_B T}\right)
\end{equation}
$f$ is the correlation factor, $g$ is the geometric factor that describes the diffusion channel connectivity, $a$ is the hop distance, $\nu$ is the pre-factor that depends on vibrational frequencies of atoms, and $E_m$ is the activation barrier associated with the atomic migration. Note that the $x$ dependence of $D$ can arise from $x$ dependence of $f$, $g$, $a$, $\nu$, and/or $E_m$. Importantly, $E_m$ is the most dominant factor that determines $D$ (and by extension $\sigma/\mu$), given that it has an exponential influence on $D$, and hence becomes the most crucial quantity to calculate/measure for rate performance estimation in any application. 

Considering reasonable battery performance metrics, such as charge/discharge at a rate of C/2 and an operating temperature of 300~K, the tolerable limit for $E_m$ in electrodes lies between 525~meV for micron-sized particles and 650~meV for nano-sized particles.\cite{rong2015materials} Minimizing $E_m$ is essential for high $\mu$, prompting efforts to identify factors that lower $E_m$. Previously proposed design principles for selecting structures with high $\mu$ include avoiding preferred coordination environments, minimizing coordination number changes during migration, and maximizing volume per anion with non-close-packed structures.\cite{rong2015materials,wang2015design} However, these principles are not universally applicable, especially for large intercalants, prompting Lu et al.\cite{lu2021searching} to refine the strategies for Ca$^{2+}$ that emphasized optimal transition state geometry not containing face-sharing polyhedra and having higher degrees of freedom. Moreover, close-packed structures (i.e., structures without a high volume per anion) have also demonstrated high $\sigma$, as with the case of Mg$^{2+}$ in spinel chalcogenides.\cite{canepa2017high,glaser2024mgb2se4,wang2019mgsc2se4,koettgen2020computational} Other structural descriptors that have been identified to correlate with $E_m$ in solids, such as spinels, garnets, and olivines, include migrating ion-anion distances,\cite{bolle2021automatic} and the `migration number'\cite{sotoudeh2022descriptor} that encompasses electronegativities, oxidation states, and ionic radii. Nevertheless, despite advancements in understanding factors that influence $\mu$ (or $E_m$) in specific systems, a generalized rule or model that is applicable across a wide variety of solid systems \cite{peng2022human} remains elusive so far.  

For models predicting $E_m$ to be practically useful, the models have to make predictions that are as accurate as and are significantly faster than experimental measurements or computational techniques. Additionally, for constructing generalizable models for $E_m$ predictions, a reliable dataset of measured or calculated $E_m$ across a wide variety of systems is necessary. In terms of measuring $E_m$, direct experimental techniques such as electrochemical impedance spectroscopy (EIS\cite{barsukov2012electrochemical, itagaki2005licoo2}), galvanostatic intermittent titration technique,\cite{kang2021galvanostatic,tang2011kinetic,sun2016high,avvaru2025alternative} and nuclear magnetic resonance based methods\cite{grey2004nmr,verhoeven2001lithium,huang2024enhanced}  are typically used. However, these techniques can be resource-intensive and contain challenges, such as (lack of) sensitivity to the short time and length scales of ionic migration, dependence on sample preparation or measurement conditions, and specific equipment requirements (e.g., inert ion-blocking electrodes in EIS). Thus, generating $E_m$ dataset(s) based solely on experimental measurements can be challenging and models that exhibit swift and accurate $E_m$ predictions can certainly be used for targeted experimentation on select materials.

On the other hand, computational strategies to estimate $E_m$ include bond valence (BV\cite{adams2006bond,brown2009recent}) analysis, density functional theory (DFT\cite{hohenberg1964inhomogeneous,kohn1965self}) based nudged elastic band (NEB\cite{jonsson1998nudged,henkelman2000climbing}) calculations and molecular dynamics (MD\cite{frenkel2002understanding,mo2012first,car1985unified}) simulations, with each technique exhibiting its own advantages and challenges. For example, BV analysis is computationally efficient but is error-prone in estimating $E_m$ due to its empirical nature involving static structures and ionic bond models.\cite{meutzner2019computational,nestler2019towards} Ab initio MD simulations enable direct estimation of $D$ but are computationally intractable for sampling dynamics in large systems and over nanosecond time scales, thereby limiting their accuracy.\cite{he2018statistical} While machine learned interatomic potentials (MLIPs\cite{deringer2019machine,chen2022universal,du2023machine}) can enable MD simulations to sample over larger length and longer time scales than ab initio MD, the potentials have to be nominally fine-tuned for specific chemistries for accuracte $D$ or $E_m$ estimation, thus resulting in higher computational costs. 

DFT-NEB calculations offer accurate and direct $E_m$ estimation by modeling the the minimum energy path (MEP) of atomic migration using an elastic band of intermediate images connected by spring forces, which is subsequently relaxed using DFT. While the accuracy of DFT-NEB can depend on the selected exchange-correlation (XC) functional, the computational cost scales with system size, and convergence to the MEP can be problematic.\cite{devi2022effect} Efforts to accelerate NEB calculations include improved path initialization \cite{smidstrup2014improved,smidstrup2014improved} and quicker energy estimation using algorithms like `Pathfinder' and `ApproxNeb',\cite{bolle2021automatic,rong2016efficient} with possible reductions in accuracy. Nevertheless, computationally demanding DFT-NEB calculations remain the state-of-the-art for accurate $E_m$ estimations in solids, making the case for models that can accurately and swiftly predict $E_m$. 

Machine learning (ML) models have been used in the recent past to understand trends in $E_m$ in specific systems. For example, Jalem et al.\cite{jalem2014efficient} developed a neural network model with features from both DFT relaxed structures and literature to predict DFT-NEB calculated $E_m$ across 72 olivine-based structures. Apart from identifying structural descriptors that primarily correlated with $E_m$, Jalem et al.'s model exhibited a validation R$^2$ score and root mean squared error (RMSE) of 0.978 and 0.0619~eV, respectively. As a follow-up work, Jalem and co-workers\cite{jalem2018bayesian} used Bayesian optimization to identify compositions with low $E_m$ ($<$0.3~eV) within the tavorite framework by training on 317 DFT-NEB calculated $E_m$, with the model exhibiting a $\sim$90\% success rate and some knowledge transfer from Li-based compositions to Na. Sendek et al.\cite{sendek2017holistic} used logistic regression on experimental $\sigma$ data available on $\sim$40 materials that were down-selected from a screening process beginning with 12,831 Li-containing compounds from the materials project (MP\cite{jain2013commentary}) and reported a X-randomization performance metric of 0.59 indicating statistical significance in predicting high $\sigma$. Kim et al.\cite{kim2022machine} investigated anti-perovskites as solid-state electrolytes by employing various ML-based regression models that were trained on $\sim$608 DFT-NEB calculated $E_m$ using 44 physical, chemical, electrical, and geometric descriptors, with their best model achieving a RMSE of 0.71~eV. A significant limitation of existing ML models is their poor generalizability due to their reliance on narrow training datasets focused on specific structures or chemistries, resulting in poor accuracy when tested out-of-domain.

An important challenge in developing ML models for materials science is the scarcity of data on critical material properties, such as $E_m$. For example, graph neural networks (GNNs) ideally require datasets with $\sim$10$^4$ data points to perform optimally,\cite{dunn2020benchmarking} while typical material science datasets often contain only a few thousand or fewer data points. Transfer learning (TL), which involves pre-training (PT) a model on a larger (material property) dataset and subsequently fine-tuning (FT) on a smaller (target property) dataset, offers a promising solution to this data insufficiency.\cite{kang2023multi} Indeed, we have demonstrated in our previous work the effectiveness of various PT/FT strategies using the atomistic line graph neural network (ALIGNN\cite{choudhary2021atomistic}) as the base architecture, showing that FT models consistently outperform models trained from scratch (i.e., without any PT) across seven diverse bulk material properties.\cite{devi2024optimal} Furthermore, we developed a multi-property pre-trained (MPT) model, which was trained simultaneously on all seven bulk material properties and exhibited better performance than pair-wise PT/FT models on a completely out-of-domain property on 2D materials.\cite{devi2024optimal} Thus, principles of TL and our constructed MPT model can be leveraged for $E_m$ predictions.

In this work, we use TL and our MPT framework to construct generalizable models for $E_m$ predictions across a wide range of structural groups, compositions, chemistries, and migration pathways. We develop four different architectural modifications for FT the MPT model, by considering either the initial and final positions of the migrating ion or an interpolated band of images that represents the migration as inputs. Additionally, we explore the utility of adding attention layers to increase the model's sensitivity to critical parameters that govern $E_m$. For training and testing, we employ a manually-curated DFT-NEB $E_m$ data obtained from literature, developed as a parallel work.\cite{devi2025Data} Our dataset contains 619 distinct migration pathways that span 58 different space groups and diverse chemistries and compositions. 

Notably, we observe our best-performing FT model, named `MODEL-3', to exhibit R$^2$ scores and mean absolute errors (MAEs) of 0.703 and 0.261~eV, respectively, on the test set. MODEL-3 outperforms both scratch GNN and classical ensemble ML models, by atleast 77.53\% on test R$^2$ scores and 17.92-44.13\% on test MAEs. Importantly, MODEL-3 displays the ability to generalize across different migration pathways, migrating ion compositions, and varying anion or transition metal chemistries within a structural framework. We find MODEL-3 to be more accurate (by 280\% on test R$^2$ and 23.24\% on test MAE) compared to $E_m$ estimated by a universal MLIP, namely the multi atomic cluster expansion (MACE\cite{batatia2025design,batatia2023foundation}). Also, as a classifier, MODEL-3 achieves an 80\% accuracy and an 82.8\% precision in identifying `good' conductors (i.e., structures with $E_m <$0.65~eV). The FT model architectures developed in this work illustrate the adaptability of the MPT model to predict other data-scarce material properties. Finally, our best-performing model should be highly useful in rapidly identifying materials with good $\mu$, which can be subsequently validated with DFT-NEB calculations or experiments, and eventually be used for batteries and other applications.   

\newpage
\section{Data description}
\begin{figure}[h!]
\centering
\includegraphics[width=0.87\textwidth]{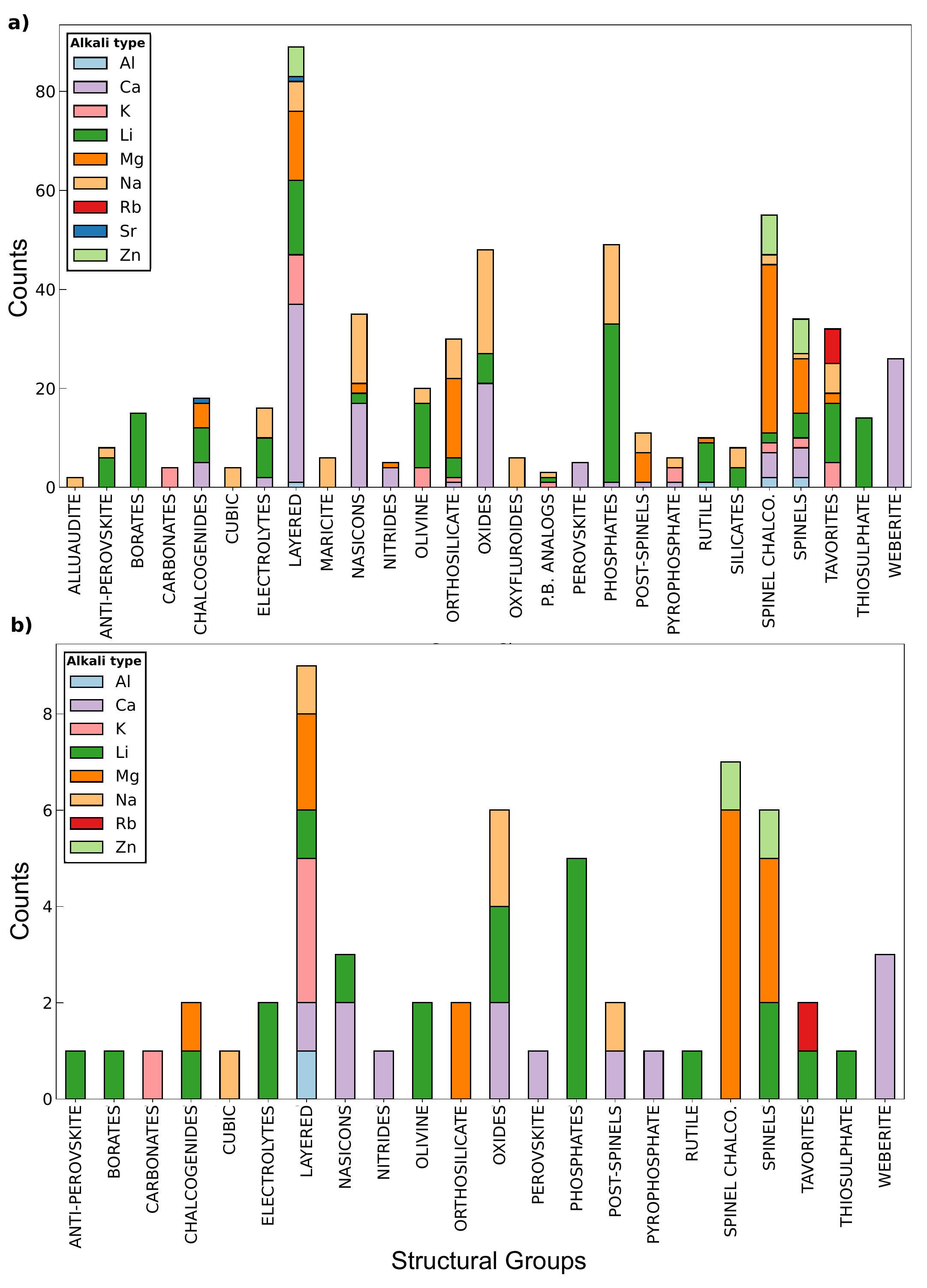}
\caption{\textbf{Data distribution of training and test dataset }. Distribution of the data in the (a) train and (b) test datasets across different structural groups in the final dataset. The colored stacked bar heights within each bar corresponds to the contribution from different intercalant ions within that structural group.}
\label{fig:data-dist}
\end{figure}

Here, we use a curated dataset comprising 619 distinct migration pathways (see Figure~{\ref{fig:data-dist}}), across various ionic compounds that have been studied as electrode or solid electrolyte materials for batteries, and the corresponding $E_m$ derived from DFT-NEB calculations. The overall dataset is available at Ref.~\citenum{devi2025Data}, as part of our additional work. The dataset encompasses both discharged (71.4\%) and charged states (23.6\%) of the electrodes, with intermediate compositions (5\%) included in select instances. Note that we refer to electrode materials with high intercalant compositions (e.g., $x \sim 1$ in Li$_x$CoO$_2$) as the corresponding discharged states, while low intercalant compositions ($x \sim 0$ in Li$_x$CoO$_2$) constitute charged states.

The majority (88.12\%) of the collected $E_m$ values were calculated using the generalised gradient approximation (GGA)\cite{perdew1996generalized} as the XC functional, with other datapoints calculated with the Hubbard \textit{U}\cite{anisimov1991band} corrected version of GGA (or GGA+\textit{U}), the strongly constrained and appropriately normed\cite{sun2015strongly} functional, and the localised density approximation.\cite{jones1989density,van1999correcting} $E_m$ values were initially gathered from published literature, and the corresponding crystal structures were obtained from the MP or the inorganic crystal structure database.\cite{hellenbrandt2004inorganic} In cases where  structures were unavailable, ground state (GS) structures were generated from appropriate parent structures using reference lattice scaling\cite{chu2018predicting} to modify the lattice parameters and/or by enumerating possible ordered configurations via the OrderDisorderedStructureTransformation class within the pymatgen package,\cite{ong2013python} and subsequent relaxations with DFT.\cite{devi2025Data} 

The dataset spans 58 distinct space groups, categorised into 27 structural groups, with migration barriers ranging from 0.03 to 8.77 eV. Prominent structural groups in our dataset include spinels, layered, olivines, tavorites, phosphates, weberites, and NaSICONs. As much as possible, we used structural groups that are quite common in the battery literature. Layered structures constitute the largest portion of the dataset, with 98 entries, followed by spinel chalcogenides, phosphates, and the general class of oxides. Other structural groups, such as alluaudites, Prussian blue analogs, and carbonates, are also represented, albeit with fewer data points. Lithium (Li)-based intercalants account for approximately 28.3\% of the dataset, followed by calcium (Ca), sodium (Na), magnesium (Mg), potassium (K), zinc (Zn), strontium (Sr), aluminum (Al), and rubidium (Rb). Further details regarding the complete dataset generation,  distribution of datapoints, and comprehensive descriptions of each datapoint can be found in Ref.~\citenum{devi2025Data}. The split of the dataset into train and test subsets for model training and evaluation and the data distribution in each subset is described in the Methods section and visualized in Figure~{\ref{fig:data-dist}}.

\section{Results}
\subsection{Architectural modifications}
Specific architectural modifications were necessary to adapt and FT our previously developed MPT model\cite{devi2024optimal} for the prediction of $E_m$. A key objective of our final trained model is that it should be able to distinguish multiple migration pathways within the same crystal structure, since that is quite typical in ionic solids. A classic example of a structure with multiple migration pathways is the layered-Li$_x$CoO$_2$, where the Li migration along Li layers is significantly faster compared to a Li migration across the Co layers. Thus, distinguishing multiple migration pathways requires that the model be provided an inductive bias on the direction of motion of the migrating ion. We provided the directional information by either considering both initial and final configurations along the migration pathway as inputs to the modified MPT model or in the form of a set of interpolated images (i.e., a band) between the initial and final images as an input. While including the initial and final configurations only accounts for the direction of motion, the band input also includes partial information on the transition state geometry. 

\begin{figure}[h!]
\centering
\includegraphics[width=0.89\linewidth]{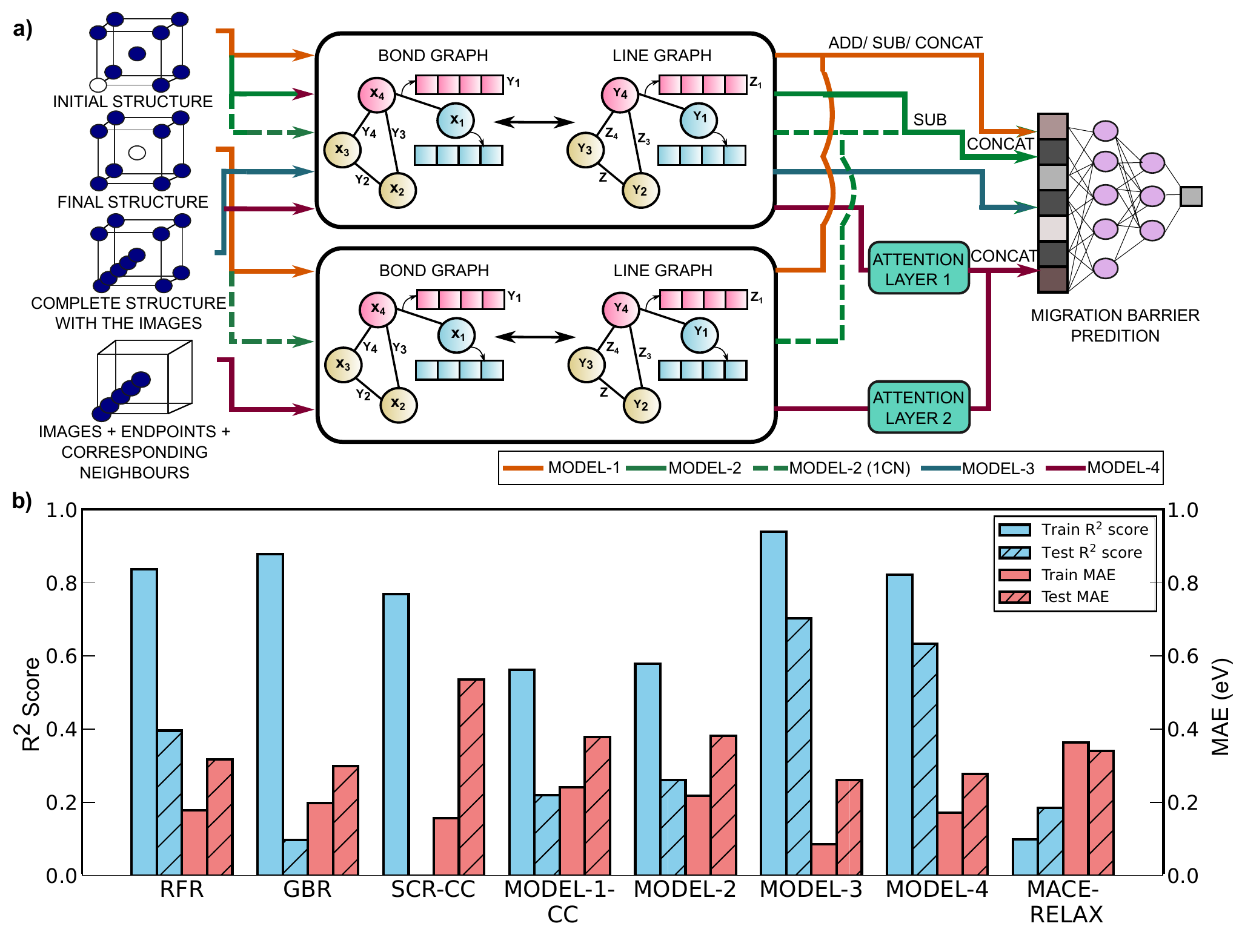}
\caption{\textbf{Modified MPT architectures used for FT and train/test scores of different models.} (a) The four different modified MPT architectures explored in this work, namely MODEL-1 (orange lines), MODEL-2 (green), MODEL-3 (blue), and MODEL-4 (magenta). Solid (dashed) input/output lines correspond to four (one) convolutions performed on the MPT model. (b) R$^{2}$scores and MAE for all models. The left (blue bars) and right (pink bars) y-axis values represent R$^{2}$ scores and MAEs (in eV). Solid and hashed bars denote the train and test scores, respectively.}
\label{fig:Models}
\end{figure}

Additionally, we varied the number of convolutions performed during FT, and also added attention-based pooling mechanisms to refine the embedding of the local geometries around the migration path. These modifications led to four different FT model architectures, namely MODEL-1 (orange lines), MODEL-2 (green), MODEL-3 (blue), and MODEL-4 (magenta), as illustrated in Figure~{\ref{fig:Models}}a. Each model uses either one or two copies of the MPT model and the solid (dashed) lines in Figure~{\ref{fig:Models}}a indicate cases where four (one) convolutions are performed during FT. Briefly, MODEL-1 includes two MPT instances and takes the initial and final configurations as inputs and pools the output of both instances (via addition, subtraction or concatenation) before $E_m$ predictions. MODEL-2 is similar to MODEL-1 but takes in a difference vector (``delta'') of embeddings between the initial and final configurations and concatenates the delta to the initial configuration embeddings. MODEL-3 is a single MPT instance that takes the interpolated band as the input. MODEL-4 builds upon MODEL-3 by adding attention blocks to two MPT instances, one that takes the full structure with the interpolated band and another that takes a sub-graph that focusses on the local environment of the band. Detailed descriptions of all FT model architectures, including scratch models, are provided in the Methods section.

\subsection{Performance of classical and scratch models}
Figure~{\ref{fig:Models}}b illustrates the R\textsuperscript{2} scores and MAEs for both training (solid bars) and test (hashed bars) datasets for different models, including classical ML models, scratch models (SCR), FT models, and DFT-NEB calculated $E_m$ using a universal MLIP (denoted as MACE-RELAX). Blue (pink) bars and the left (right) y-axis values represent the R\textsuperscript{2} scores (MAEs in eV). For models with multiple scenarios, the best-performing model scores are selected for the plot. For example, MODEL-1 has three different scenarios depending on whether the final set of embeddings from the two MPT instances are added, subtracted, or concatenated (CC) and Figure~{\ref{fig:Models}}b displays the scores for the MODEL-1-CC, which exhibits the best performance. The magnitude of the scores for all the models are tabulated in Table~S5 of the supporting information (SI). Note that large negative R\textsuperscript{2} scores (such as -1.241 for SCR-CC) are denoted as zeros for illustrative purpose. A model's performance is considered good if it achieves both a high R\textsuperscript{2} score and a low MAE. 

Among the classical ML approaches, namely random forest regression (RFR) and gradient boosted regression (GBR), we observe RFR to outperform GBR on the test data even though GBR has better training scores, indicating that the GBR model was likely overfit. Importantly, our scratch models (SCR-CC in Figure~{\ref{fig:Models}}b) perform inferior in comparison to the classical models in terms of both training and test scores. Among the scratch models, the model where the embeddings of the initial and final configurations were subtracted from each other (SCRATCH-SUB) exhibited the worst performance with highly negative R\textsuperscript{2} scores, which is mainly due to the components of the embedding vector tending to zero after subtraction, thus failing to capture the small differences between the initial and final configurations. On the other hand, adding or concatenating the embedding vectors in scratch models offered better performance than SCRATCH-SUB. 

\subsection{Performance of TL models}
Comparing the SCR-CC model with the FT MODEL-1 architecture, we find that TL does help in improving the performance of the graph-based model compared to scratch, with MODEL-1-CC showing a 29.3\% decrease in test MAE compared to SCR-CC. However, MODEL-1-CC's performance is still poorer in comparison to classical ML models, with a 19.2\% higher test MAE than RFR (Figure~{\ref{fig:Models}}b). The likely reasons for the inferior performance of MODEL-1-CC is its poor training scores, which can be attributed to MODEL-1-CC's failure to distinguish multiple pathways within the same structure and eventually predicting identical $E_m$ for different paths. Specifically, MODEL-1-CC failed to distinguish the distinct migration pathways in $\sim$~81.1\% of systems that had multiple pathways in our training dataset. Thus, MODEL-1-CC is unable to utilize the initial and final configuration inputs to infer differences in the direction of migration. To verify that the poor learning of multiple pathways is the main bottleneck in the performance of MODEL-1-CC, we trained a variant of the model, MODEL-1-CC-SINGLE-PATH, using only a single migration pathway per structure in our train/test datasets. As anticipated, removing multiple pathways resulted in significantly improved training scores and enhanced test performance (test R\textsuperscript{2}: 0.603, test MAE: 0.249, see Table~S5). Thus, further modifications to the graph network architecture, going beyond MODEL-1-CC, are needed if models are to distinguish multiple pathways within the same structure.  

To better capture the distinctions among multiple migration pathways, we modified the MODEL-1 architecture to generate MODEL-2, which emphasized the differences between the initial and final configurations via the calculation of the delta vector. Indeed, the proportion of datasets with multiple migration pathways that are not classified as distinct reduced to 71.1\% within the training dataset with MODEL-2 compared to MODEL-1 (81.1\%). However, MODEL-2 exhibited only an improvement of 3.02\% and 19.18\% in terms of training and test R$^2$ scores, with a corresponding increase in test MAE of 0.8\%, compared to MODEL-1. Importantly, MODEL-2's performance remains poorer than classical ML models, suggesting that even better modifications to the graph network are necessary.

To improve the ability of the graph network to identify different migration directions, we generated MODEL-3, which is a standard MPT model architecture but with the input structures augmented with positional data from three interpolated images alongside the initial and final sites. Including the interpolated images creates a band representation, or a guess of the pathway along which the ion is likely to migrate, with some information of the transition state. Including the band representation significantly improved the performance of the MPT model after FT, with MODEL-3 posting a 77.52\% increase in test R$^2$ score (0.703) and a 17.92\% decrease in test MAE (0.261~eV) compared to the RFR model (Figure~{\ref{fig:Models}}b). Thus, incorporating information on the band provides a clear intuitive bias to the graph model on the direction of motion, and the model is able to distinguish different pathways within the same structure, resulting in a clear performance improvement on the test set. Notably, only 27.2\% of the training datapoints corresponding to multiple migration pathways had their absolute errors greater than 0.1~eV with MODEL-3. 

In an effort to enhance the emphasis on interpolated images, we integrated attention layers, a fundamental component of transformer architectures widely utilised in large language models \cite{zhang2023turing,vaswani2017attention} and created MODEL-4. Attention blocks facilitate the creation of context-aware embeddings, significantly improving tasks such as text generation and translation in language models. We adapted the attention architecture by employing atom embeddings as inputs to generate query, key, and value vectors, thereby enabling an understanding of atomic interactions or the ``attention'' between atoms within the structure. In principle, including attention should heighten the model's sensitivity to critical features that influence $E_m$ predictions, since MODEL-4 leverages embeddings obtained directly from the attention layers. Indeed, MODEL-4 achieves superior test R\textsuperscript{2} scores and MAEs compared to RFR (Figure~{\ref{fig:Models}}b) and is able to distinguish multiple migration pathways within the same structure, similar to MODEL-3. However, MODEL-4's test R$^2$ scores are $\sim$9.82\% lower and its train MAEs are $\sim$101.18\% higher than those of MODEL-3. The test MAE values for MODEL-3 and MODEL-4 are quite similar (6.13\% deviation). Thus, we find that adding attention layers to MODEL-3 is not necessarily a helpful addition to improve performance, with MODEL-3 being the best FT model we have so far. 

Given that our dataset is new, we benchmark our models against a universal MLIP (MACE-MP-0\cite{batatia2025design,batatia2023foundation}), which was used as the energy and force evaluator in NEB calculations (Figure~{\ref{fig:Models}}b). Since the initial and final configurations in our dataset were predominantly unrelaxed structures (with DFT), we considered NEB calculations with MACE, both including initial/final configuration relaxation (MACE-RELAX in Figure~{\ref{fig:Models}}b) and excluding any relaxation. Unsurprisingly, the MACE-NEB calculations with initial/final configuration relaxation significantly outperformed the calculations without any relaxations, with R$^2$ scores and MAEs of 0.185 and 0.340~eV on the test dataset, respectively. Despite a test MAE that is only 30\% higher than MODEL-3, MACE-RELAX had a notably lower R² score, indicating its inability to capture qualitative trends in $E_m$ effectively. Additionally, while MACE-RELAX is able to distinguish different pathways in a structure, the absolute errors of MACE-RELAX predictions in such cases are significantly higher than MODEL-3. For example, in Na-orthosilicates (Na$_2$FeSiO$_4$) and Ca-weberites (Ca$_{1.5}$Ni$_2$F$_7$), the absolute errors in identifying multiple pathways varied between 1-2.5 eV with MACE-RELAX compared to 0.003-0.17~eV with MODEL-3. Thus, we find our FT MODEL-3 to be better at both quantitative and qualitative predictions of $E_m$ compared to MACE-RELAX. This analysis also underscores the importance of using both R\textsuperscript{2} and MAE metrics to comprehensively assess a model's performance. 

\subsection{Predictions vs.\ target}
\begin{figure}[h!]
\centering
\includegraphics[width=\linewidth]{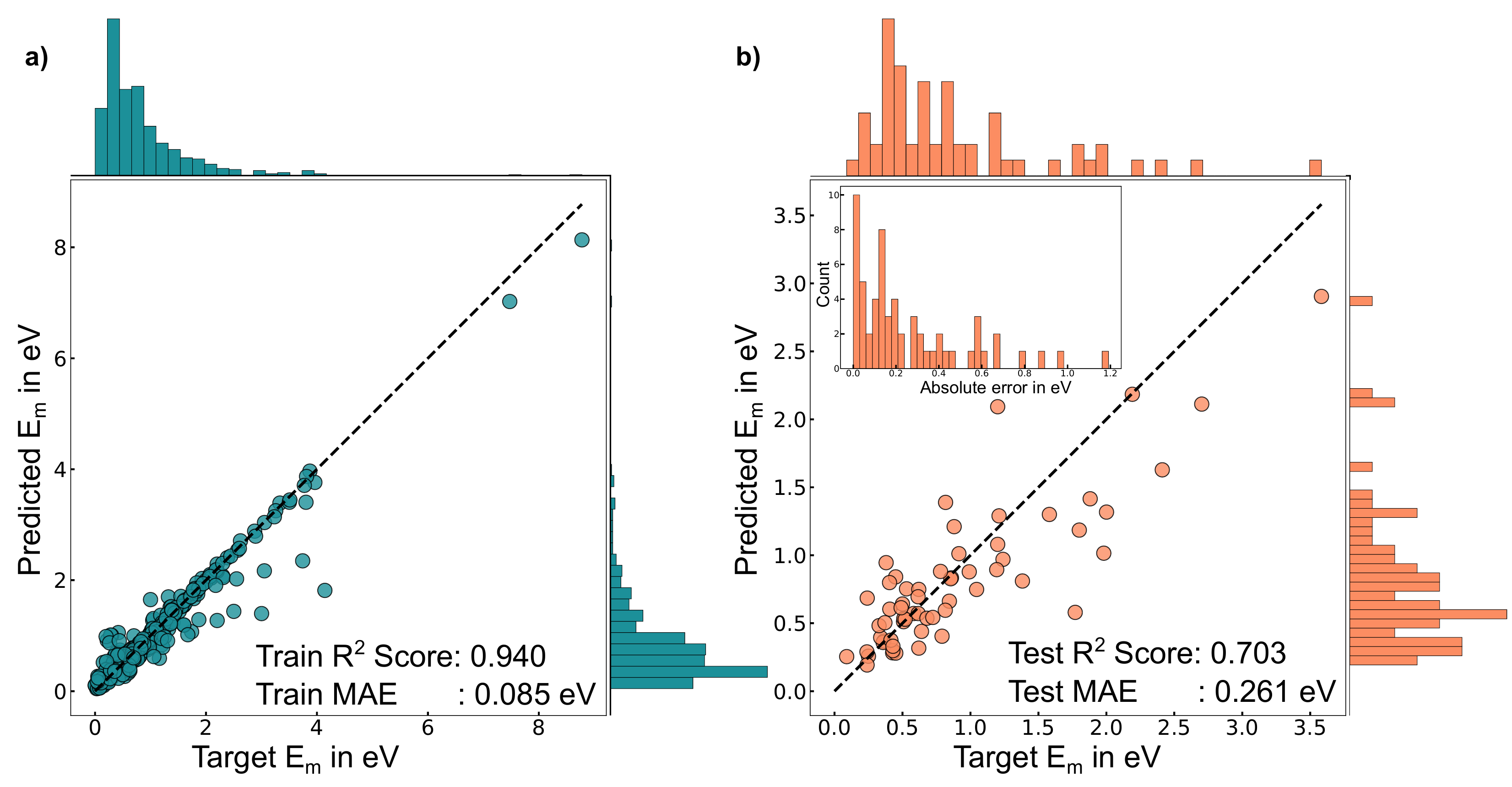}
\caption{\textbf{Prediction vs. target $E_m$.} Parity plots for a) training and b) test datasets, respectively. The inset histogram in panel b shows the frequency distribution of the absolute error range in eV. The histograms on the right and top margins illustrate the frequency distribution in the prediction and target $E_m$ in both datasets.}
\label{fig:pred_vs_target}
\end{figure}

Panels a and b of Figure~{\ref{fig:pred_vs_target}} depict the comparison between the predicted and the target $E_m$ values for the training and test datasets, respectively, for our best-performing model, MODEL-3. The prediction versus target parity plots for other models are compiled in Figures~S2 and S3 of the SI. The accompanying histograms along the x- and y-axis in Figure~{\ref{fig:pred_vs_target}} margins show the distributions of target and predicted $E_m$ values. Both training and test R\textsuperscript{2} scores and MAEs are indicated within the panels. Although the frequency of under- and over-predictions with MODEL-3 in the training set is similar, under-predictions tend to have larger absolute errors compared to over-predictions. Specifically, 12 out of 19 datapoints with absolute errors greater than 0.5 eV are under-predicted by MODEL-3, averaging an error of approximately 0.99 eV. The highest under-prediction error occurs in oxide-CaCu\textsubscript{2}O\textsubscript{3}, with a predicted $E_m$ of 1.82 eV against a target of 4.14 eV. Notably, 11 of these 12 under-predicted datapoints have target $E_m$ exceeding 1.5 eV, except for LiFeBO\textsubscript{3} (space group: P21/c, path: 4) with a target $E_m$ of 1.16 eV. Among the 12 under-predictions, five cases involve oxides and borates.

The inset histogram in Figure~{\ref{fig:pred_vs_target}}b illustrates the frequency distribution of absolute errors in the test dataset by MODEL-3. Importantly, $\sim$32\% of the test datapoints predicted by MODEL-3 have an absolute error below 0.1 eV, representing high accuracy in the predictions of MODEL-3. Additionally, 38\% fall within a moderate accuracy range (0.1 to 0.3 eV absolute error), while 30\% exceed an absolute error of 0.3 eV, indicating low accuracy. The lowest absolute error ($<$~0.1~eV) in the test dataset is observed in the case of spinels and spinel chalcogenides, which contribute about 16\% of the training dataset and contain fairly typical tetrahedral-octahedral-tetrahedral migration pathways.\cite{sai2017influence} Of the 18 test datapoints with absolute errors greater than 0.3~eV, the average MAE across the 18 datapoints is $\sim$0.60~eV with 11 $E_m$ values being under-estimated by more than 0.5~eV by MODEL-3. The highest absolute error in this group is observed for KNiF\textsubscript{3}, with a target $E_m$ of 1.77~eV and an under-prediction error of 1.19~eV, which may be due to the high levels of distortions that the perovskite NiF$_3$ is subjected to as K$^+$ moves across the structure and not captured in our model. 

The frequency of under- and over-predictions of $E_m$ across the four FT models is similar (Figure~S2), except MODEL-2 that exhibits about 20\% more instances of over-prediction on average during training. Approximately 57\% and 54.7\% of training samples from MODEL-1 and MODEL-2, respectively, achieve an absolute error below 0.1~eV. In comparison, MODEL-3 and MODEL-4 achieve significantly higher accuracy, with 81.6\% and 58.3\% of their training samples that have an absolute error below 0.1~eV. Additionally, samples with absolute errors equal to or exceeding 0.5~eV constitute around 12\% for MODEL-1 and 10\% for MODEL-2, 3\% for MODEL-3, and 6\% for MODEL-4, which explains the superior training MAE observed in MODEL-3 (Figure~{\ref{fig:pred_vs_target}}a). While these training statistics may suggest that MODEL-3 and MODEL-4 are overfitting to the training dataset, the fact that both models exhibit superior performance on the test set (Figure~{\ref{fig:pred_vs_target}}b and Figure~S2f) compared to MODEL-1 and MODEL-2 indicate that models with band inputs do generalize better to predict $E_m$ compared to models that take only the initial and final configurations as input.

\subsection{Distinguishing migration pathways}
\begin{figure}[h!]
\centering
\includegraphics[width=\linewidth]{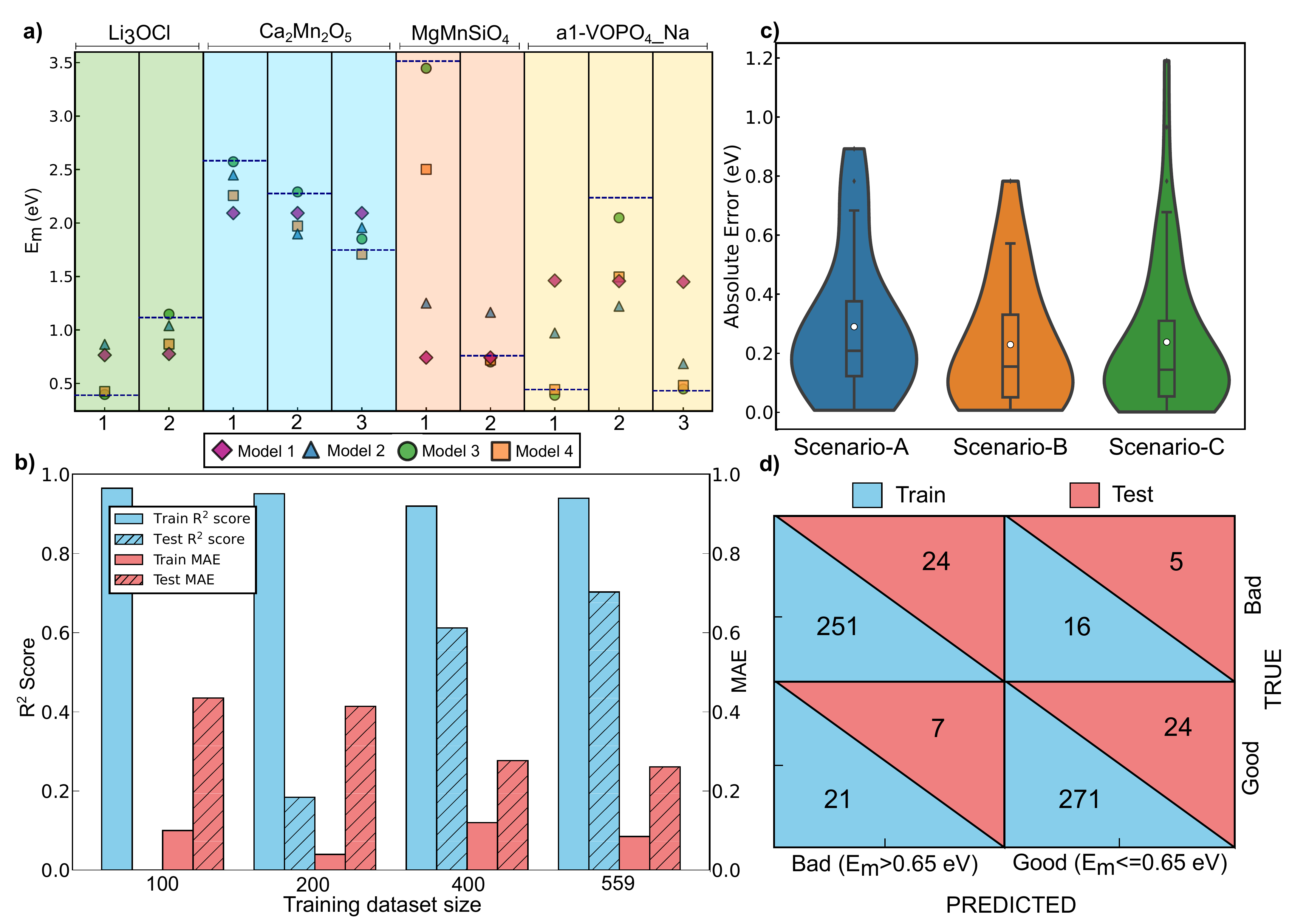}
\caption{\textbf{Analysis of model performance.} a) Race-track plot illustrating the predictions made by the four different FT models (represented by distinct colored symbols) for different pathways in a given structure. `a1' refers to a specific VOPO$_4$ polymorph of space group P4/n. Horizontal lines are the target $E_m$ values for each path. b) Bar chart showing the train and test scores of MODEL-3 with different FT train dataset sizes. c) Violin plot signifying the distribution of absolute errors among the test datapoints under scenarios A (blue), B (orange), and C (green). d) Confusion matrix quantifying the classification performance of MODEL-3 in the train (upper blue triangles) and test (lower pink) datasets.}
\label{fig:analysis}
\end{figure}

The ability of different models to distinguish multiple migration pathways within the same structure is illustrated using race-track plots in Figure~{\ref{fig:analysis}}a, which displays predictions made among training datapoints. We selected representative systems (Li$_3$OCl, Ca$_2$Mn$_2$O$_5$, MgMnSiO$_4$, and VOPO$_4$, top x-axis of Figure~{\ref{fig:analysis}}a) featuring two or three distinct migration pathways (bottom x-axis) and involving four different intercalants (Li, Ca, Mg, and Na) that are major contributions to the dataset for this analysis. The y-axis in Figure~{\ref{fig:analysis}}a represents the $E_m$ values in eV, with target values highlighted by horizontal black dashed lines within each track. Note that the VOPO$_4$ structure considered for this analysis is its charged state with Na as the intercalant - hence the notation 'VOPO$_4$\_Na'. The tracks are color-coded to differentiate among the intercalants, with the FT models indicated by different colored symbols.

Data from Figure~{\ref{fig:analysis}}a indicate that MODEL-1 (pink diamonds) consistently predicts identical $E_m$ across different pathways within the same structure, for all systems, indicating MODEL-1's inability to distinguish different paths. In contrast, MODEL-2 (blue triangles) shows variations in predicted $E_m$ across different pathways, albeit with substantial errors in its predicted $E_m$ versus the target. Thus, the inductive bias introduced in MODEL-2 is able to distinguish pathways compared to MODEL-1 for the systems in Figure~{\ref{fig:analysis}}a. Similarly, MODEL-3 (green circles in Figure~{\ref{fig:analysis}}a) and MODEL-4 (orange squares) are able to distinguish different pathways as well, indicating that band inputs are useful as tools to add inductive bias with respect to the direction of migration. While MODEL-3 and MODEL-4 exhibit predicted $E_m$ values that are closer to the target values, the data points in Figure~{\ref{fig:analysis}}a are within the training set and hence not a direct reflection of the generalizability of these models. 

\subsection{Influence of dataset size}
To quantify the influence of dataset size during FT on the prediction accuracies for the best-performing MODEL-3, we plot the train and test R$^2$ scores and MAEs for three lower FT dataset sizes, namely 100, 200, and 400 in Figure~{\ref{fig:analysis}}b. Note that 559 is the maximum training set size of our $E_m$ dataset, while the test dataset remained the same for all training dataset sizes. The notations used in the figure are identical to Figure~{\ref{fig:Models}}b. As expected, we observe significant improvement in the test scores (both R$^2$ and MAE) as the training dataset size increases. Also, the model that uses 559 training datapoints offers the best train as well as test scores. The significant disparity in the train and test scores for models trained with lower dataset sizes (especially at 100 and 200 datapoints in Figure~{\ref{fig:analysis}}b) indicates the tendency of the model architecture to overfit at very small train dataset sizes. 

\subsection{Generalization across pathway, composition, and chemistry}
To better understand the generalization abilities of MODEL-3 and its possible use as a screening tool for battery applications and beyond, we analyze its performance over three different scenarios that are encountered in the train/test datasets (see Figure~{\ref{fig:analysis}}c). Scenario A involves structures where one migration pathway is in the test set, while the remaining pathway(s) are in the train set, quantifying the model's ability to generalize across different migration pathways. Scenario B consists of systems where the charged (or discharged) composition appears in the train set, while the corresponding discharged (charged) state appears in the test set, signifying the model's ability to generalize across compositions of an intercalant within a host framework. Scenario C determines the model's generalizability across different chemistries since it encompasses instances where either the intercalating ion or the host structure's anion/cation in the test set differs from what the model has seen in the train set. 

Figure~{\ref{fig:analysis}}c plots the absolute error (in eV) in the $E_m$ predictions on the test set for the three scenarios as violins. The lower and top edges of the violins correspond to the range of the absolute error for each scenario. The empty circle is the mean, and the solid black line is the median of the distribution. Scenarios B and C have similar mean ($\sim$0.23~eV) and median ($\sim$0.14~eV) absolute errors that are lower than Scenario A ($\sim$0.29 and 0.20~eV), and have a larger number of data points with low absolute errors ($<0.1$~eV). This indicates that MODEL-3 generalizes better across intercalant composition and chemistry compared to migration pathways. Additionally, prediction confidence is highest for scenario C (followed by scenarios B and A), as 68\% of systems in scenario C have a test MAE of 0.11~eV (i.e., in the high/moderate accuracy range). Thus, our analysis suggests that MODEL-3 can offer robust predictions for a given system when it has seen structurally (not necessarily compositionally or chemically) similar systems during training, which can be quite useful when used as a screening tool. Importantly, our analysis indicates that distinguishing directionality of migration within the same structure is perhaps the hardest task for the graph network architectures considered in this work, highlighting the role of difficult-to-describe local coordination environments in differentiating $E_m$. 

\subsection{Classification metrics}
To examine if MODEL-3 can be used as a screening tool to classify structures as good ($E_m \leq 0.65 eV$) or bad ($E_m$ > 0.65 eV) ionic conductors, instead of being used as a 'regression' tool to predict absolute $E_m$, we segregated the datapoints that fall into the above two categories in both the train and test sets. Figure~{\ref{fig:analysis}}d illustrates the confusion matrix, which tabulates MODEL-3's performance in identifying a good or a bad conductor in the train (lower blue triangles) and the test (upper pink) sets, respectively. The overall accuracy of the test (train) predictions is 80.0\% (93.4\%), which represents the number of correct predictions in the overall test (train) samples, highlighting the potential use of MODEL-3 as a classifier of structures as good/bad conductors. The sensitivity (recall) or the ability of the model to classify actual good and bad conductors in the test set is 77.4\% and 82.8\%, respectively. Notably, the precision in classifying good conductors is higher (82.8\%) than that of bad conductors (77.4\%), implying that statistically MODEL-3 is marginally less likely to falsely classify a structure to be a good conductor than a bad conductor. Note that the threshold criterion in identifying good and bad conductors in this analysis ($E_m$=0.65~eV) is arbitrary, and using a different threshold will modify the confusion matrix.

\section{Discussion}
For the accurate estimation of $E_m$ in ionic solids, we present a MPT model that has been modified and FT specifically for precise $E_m$ predictions over a wide range of crystal structures and migration pathways. Via modifications to the pre-trained MPT model to introduce inductive bias on the directionality of a migration path, we FT four different MPT model architectures on a curated dataset of 619 DFT-NEB-calculated $E_m$ values obtained from literature. Importantly, we find our MODEL-3 architecture, that takes a band input, to be the best performing, with test R$^2$ and MAE of 0.703 and 0.261~eV, respectively. We observe MODEL-3 to not only distinguish multiple pathways in a given structure, but also generalize well across intercalant chemistries and compositions. Furthermore, we find that MODEL-3 can be used as a classification tool, with an accuracy rate of 80\% on classifying a structure as a good ($E_m \leq 0.65$~eV) or a bad ($E_m > 0.65$~eV) conductor. Thus, our best-performing model should be a useful tool in the screening of materials with high $\mu$ for battery applications and beyond.

The performance of FT models can be further improved by including a larger number of data points in our dataset, which will require more systematic DFT-NEB calculations. Note that we expect our FT models to outperform scratch models with additional data, given that the FT models that are trained with fewer data points ($\sim$200, Figure~{\ref{fig:analysis}}b) exhibit similar performance to scratch models trained on the full dataset (Figure~{\ref{fig:Models}}b), identical to our observations in our previous work.\cite{devi2024optimal} In addition to presenting MODEL-3 for efficient $E_m$ predictions, our study highlights possible strategies to modify the MPT model for targeted FT on specific datasets. Our proposed modifications (in Figure~{\ref{fig:Models}}a) allow (partially) capturing global and local structural details that correlate directly with our target property ($E_m$). Thus, similar modifications of our MPT model, with careful hyperparameter optimization, can be used for FT on different target properties (such as adsorption energies on surfaces, point defect formation energies, etc.\cite{witman2023defect}) that are difficult to calculate/measure.

The predictive accuracy of our best-performing MODEL-3 still remains considerably lower than the R$^2$ scores and MAEs reported in our previous work\cite{devi2024optimal} for `simpler' property predictions, such as formation energy (R$^2$ and MAE of 0.774 and 0.089 eV). We attribute this lower accuracy to three factors. First, the model may have inadequate level of inductive bias resulting in insufficient learning of both local and global structural features that determine $E_m$. Second, the dataset may contain intrinsic noise, since all $E_m$ have not been calculated at the same level of theory and ensuring all $E_m$ are at the same level of theory is beyond the scope of this work. Third, ALIGNN only considers bond and angle embeddings and using more advanced GNN architectures that consider many-body interactions may improve prediction accuracy.

We observe that MODEL-3 seems to generalize more efficiently across intercalant compositions and chemistries than it does across different migration pathways within a specific structure (Figure~{\ref{fig:analysis}}c), which can be attributed to the following factors. First, the underlying migration pathway can be identical (or similar) across different material compositions and chemistries. For instance, spinel structures exhibit a consistent tetrahedral-octahedral-tetrahedral migration path regardless of whether the composition is charged or discharged. Also, the migration path in spinels remains the same irrespective of variations in the migrating ion, transition metal cation, or the anion. Thus, if a model learns the key factors contributing to the $E_m$ in one spinel structure, it may be able to generalize well on other spinels. Indeed, we find MODEL-3's prediction accuracy to be high for spinels (absolute errors in $E_m <$0.1~eV). Second, the magnitude of $E_m$ can be determined by local structural `motifs' that define the migration pathway. While global structural features like lattice parameters, composition, and transition metal/anion identity determine the overall potential energy landscape, local structural features (such as coordination number changes, bond distances, rotation or distortion of polyhedral units) dictate the 'local' energetics near a saddle point. Thus, limitations in the model's ability to capture the importance of such local features may affect its generalization ability.

Surprisingly, the addition of attention blocks into MODEL-3 (resulting in MODEL-4, Figure~{\ref{fig:Models}}a) did not yield a significant enhancement in the predictive accuracy (Figure~{\ref{fig:Models}}b). Our rationale for the inclusion of attention blocks was to provide the model with a better understanding of the ``context" or importance of an atom or a local structural motif along the migration pathway. This is why we employed the dual-input approach, where we used the overall band and the sub-graph of the band's local coordination environment. One possible reason for the ineffectiveness of attention blocks is the limited size of our dataset. Indeed, attention layers have been demonstrated to be highly effective in large language models trained on billions of tokens,\cite{derose2020attention} while we only have a total of 619 data points. Nevertheless, as the scientific community contributes and makes the $E_m$ dataset more general, it will be worthwhile to revisit the utility of attention blocks in extracting atom-level importance in material property predictions.

MACE-NEB-RELAX exhibited convergence difficulties in 68 systems, which may be potentially resolved by adjusting BFGS steps, interpolation methods, or other hyperparameters. In certain cases, convergence issues observed in MACE-NEB-NO-RELAX were resolved upon relaxing the initial and final images. Across the complete dataset, MACE-NEB-RELAX achieved an overall R\textsuperscript{2} score of 0.10 and an MAE of 0.36 eV, emphasizing MODEL-3’s superior accuracy. Despite its low R\textsuperscript{2} score, MACE-NEB-RELAX provided accurate $E_m$ estimations and reasonable geometries for intermediate images upon a NEB calculation, particularly for `simple' systems containing single anions  such as spinel oxides, spinel chalcogenides, post-spinels, and layered structures. If convergence challenges can be mitigated, output geometries from MACE-NEB-RELAX can serve as initial guesses for DFT-NEB calculations, which may reduce computational costs.

\section{Conclusion}
In conclusion, precise $E_m$ estimation is paramount for quantifying ionic conductivity, a material property that is crucial in selecting materials for batteries and other applications. Building upon our previous work involving simultaneous PT of a graph-based neural network on seven bulk material properties, we introduced four modifications into a MPT model to build inductive bias on distinct migration pathways within a structure. Subsequently, we FT each model using a curated dataset of 619 DFT-NEB-calculated $E_m$ values sourced from literature for precise $E_m$ predictions across various migration pathways, structural frameworks, and intercalant chemistries. MODEL-3, which involved a single MPT model instance that took a band configuration as input, demonstrated the best performance, achieving a R\textsuperscript{2} score of 0.703 and a MAE of 0.261~eV on the test set and successfully identified multiple migration pathways within a structure. Additionally, MODEL-3 showed an ability to generalize well across intercalant compositions and chemistries. Furthermore, MODEL-3 effectively classified good ($E_m \leq 0.65$~eV) and bad ($E_m > 0.65$~eV) ionic conductors, achieving 82.8\% precision and 77.4\% sensitivity for identifying good conductors, highlighting its possible use in high-throughput screening approaches to identify novel materials. Beyond improved $E_m$ predictions, our work demonstrates effective strategies for adapting and modifying PT graph architectures to FT on material properties that are scarcely available, thereby addressing a critical challenge of data insufficiency for ML in materials science.

\section{Methods}
\subsection{Data processing}
The dataset generated from Ref.~\citenum{devi2025Data} was split into the training and test set in the ratio, 90:10. Our final test set consists of 60 datapoints across 21 distinct structural groups, with the training set consisting of 559 datapoints across 27 groups. As established in our previous work,\cite{devi2024optimal} it is important to standardise and normalise the target metric when we PT/FT on different properties of varying scales and units. Thus, we standardized and normalized the training and the test datasets using the training data statistics, namely, the minimum, the maximum, the mean, and the standard deviation. 

Panels a and b of Figure~{\ref{fig:data-dist}} show the stacked bar charts illustrating the data distribution across the structural groups in the training and the test datasets, respectively. The length of each color-coded stacked bar in each panel of Figure~{\ref{fig:data-dist}} represents a unique intercalant and is equal to the number of datapoints with the corresponding intercalant in that structural group. Our goal in splitting the train and test subsets was to generate a structurally diverse test set with a focus on the most prevalent structural groups in the training data, which ensures that model performance is mainly assessed on groups with sufficient training data while also including structural diversity. We carefully constructed the test dataset by using the following strategies: 

\begin{enumerate}
    \item Quasi-weighted distribution: The data distribution in the test set is similar to the  distribution of structural groups found in the training set, ensuring a balanced evaluation. 
    \item Structural groups constituting between 1-2\% of the complete (train+test) dataset were represented by a single datapoint in the test set. This prevents the model from being unfairly penalized due to limited exposure to these groups during training. Examples of structural groups identified in this step include rutile, pyrophosphate, perovskite, nitride, anti-perovskite, borates, and carbonates. 
    \item The test set excluded structural groups with limited representation in the overall dataset (i.e., $<$1\%). We also excluded structural groups from the test set with similar chemical compositions to those already present in the test set but with different space groups, even if their contribution to the overall dataset is slightly $>$1\%. For example, we excluded silicates that comprise $\sim$1.2\% of the overall dataset from the test set, because we already included orthosilicates with similar composition but different space groups in the test set. This strategy prioritizes more abundant and structurally distinct groups in the test dataset, thus providing a robust evaluation of our models. 
    \item Random sampling within groups: Once the desired percentage contribution for each structural group was determined, datapoints were randomly selected from the pool of available data to represent that group in the test set. 
\end{enumerate}

\subsection{Graph network model and pre-training}
GNNs are well-suited for capturing the inherent atomic connectivity within crystal structures, leading to improved property predictions. We employed the ALIGNN architecture, a GNN variant, due to its demonstrated ability to enhance accuracy of material property predictions and to generalize on out-of-distribution tasks.\cite{omee2024structure,devi2024optimal} The ALIGNN architecture, typically comprising seven layers, processes atom (X), bond (Y), and bond-angle embeddings (Z) in layers 1, 2, and 3, respectively. The X, Y, and Z embeddings serve as input to layers 4 and 5, which perform edge-gated graph convolutions (E-GGC) multiple times.\cite{dwivedi2023benchmarking} The outputs from layers 4 and 5 (typically referred to as the ALIGNN layers) are pooled and subsequently passed through a multi-layer perceptron (MLP) to generate a property prediction. Further details regarding the construction of GNNs and the ALIGNN architecture are available in the literature.\cite{choudhary2021atomistic,xie2018crystal, duval2023hitchhiker} The default ALIGNN architecture that we employed and the associated hyperparameters are compiled in Section~S1 and Table~S1 of the SI.

The generalised MPT model, as developed in our previous work,\cite{devi2024optimal} was trained on a comprehensive multi-property dataset. Specifically, the MPT model was constructed by modifying the final MLP layer of the ALIGNN architecture to include seven prediction heads, each corresponding to one of the seven material properties, as illustrated in Figure~S1 of the SI. Each data point (structure) is mapped to a one-hot encoded vector and a property vector of dimension 7, where the former indicates the availability of a specific property for the structure and the latter contains the corresponding property values. We modified the loss function, as in Equation~{\ref{eq:loss}}, where $y_p$ and $y_t$ represent the predicted and target values, $i$ is the property index, $N$ is the number of properties, and $\delta^i$ is the entry of the one-hot encoded vector for property $i$. This MPT model, which was trained simultaneously on all seven properties,\cite{devi2024optimal} was utilised as a PT model for FT on the target property ($E_m$) in our work. 
\begin{equation}
\label{eq:loss}
\mathcal{L} = \frac{1}{N}\sum_{i=1}^{N} |y_p^i - y_t^i|\delta^i
\end{equation}

\subsection{Fine-tuning model architectures}
Figure~{\ref{fig:Models}}a illustrates the four model architectures explored in this work, which are explained in detail below. We designed a model architecture incorporating either one or two copies of the pre-trained MPT model, since a given MPT (or ALIGNN) model can take only one structure as an input. Each copy received either the initial or final or an interpolated band configuration of the migration pathway as input. During FT, the model was allowed to re-train fully on the $E_m$ dataset (i.e., all weights and biases were allowed to be changed), with model parameter initializations coming from the MPT model, based on our observations of optimal FT strategies in our previous work.\cite{devi2024optimal} The outputs from the (average) pooling or the attention layers of the MPT model(s) corresponding to the initial/final/band configurations were combined using various strategies (as explained below) before being passed to an MLP for the final property ($E_m$) prediction. Thus, we constructed four distinct model configurations for learning $E_m$, and the associated hyperparameters for each configuration is provided in Section~S1 and Table~S1 of the SI.

\subsubsection{MODEL-1: Pooled Embedding Combination} 
In this model architecture (follow orange lines in Figure~{\ref{fig:Models}a}), we initialized two instances of the MPT model, with one instance receiving the initial configuration of the migration pathway as its input and another receiving the final configuration. Following four E-GGC operations in the ALIGNN layers and subsequent pooling, the resulting embeddings from the two MPT model instances were either concatenated (CONCAT or CC), added (ADD), or subtracted (SUB) from each other before passing to a fully connected MLP with two hidden layers for $E_m$ prediction. For illustrative purposes, empty circles within the cubic structures represent vacant sites in Figure~{\ref{fig:Models}}a.

\subsubsection{MODEL-2: Delta Vector Concatenation} 
Similar to MODEL-1, this model also uses two instances of the MPT model and is represented by the green lines in Figure~{\ref{fig:Models}}a. The first embedding vector in this model was the pooled output following four E-GGC operations in the MPT instance that received the initial configuration as input. The second embedding vector, which we term the difference vector or delta, was calculated by subtracting the pooled vectors from two MPT models, with one instance receiving the initial configuration and the other instance the final configuration. Note that for calculating delta, the pooled vectors were obtained after one E-GGC operation instead of the standard four (as highlighted by dashed green lines instead of solid green lines in Figure~{\ref{fig:Models}}a). Subsequently, the delta vector was concatenated with the first embedding vector and the resultant vector passed to the final MLP for $E_m$ prediction.

\subsubsection{MODEL-3: Interpolated Images} 
MODEL-3 is perhaps our simplest model that uses only one instance of the pre-trained MPT model and is indicated by blue lines in Figure~{\ref{fig:Models}}a. The input to the MPT model in this architecture is the migration pathway, which we generate by a linear interpolation of three images between the initial and final atomic configurations. Note that linear interpolation is the typical technique utilized to initialize the MEP in DFT-based NEB calculations. Thus, the input to the MPT model is a single structure with initial, final, and three interpolated image sites being occupied by the migrating ion, with the resultant pooled vector (after four E-GGC operations) passed to the MLP for $E_m$ prediction. 

\subsubsection{MODEL-4: Interpolated Images with Attention} 
MODEL-4 builds upon MODEL-3 by incorporating attention blocks to generate a refined pooled representation of the ALIGNN outputs, as represented by the magenta lines in Figure~{\ref{fig:Models}}a. The objective of the attention-based pooling is to identify and prioritize the most influential nodes (atoms) of the structure for the target variable ($E_m$). To calculate the attention scores ($A$), we learn projection matrices that map the graph's atom embedding vectors ($X$) to its corresponding query ($Q$), key ($K$), and value ($V$) representations, via learnable weights ($W$). ReLU stands for the rectified linear unit activation function. 
\begin{align*}
Q, K, V &= \text{ReLU}(XW_{Q,K,V})
\end{align*}
$Q$ and $K$ are subject to batch matrix multiplication followed by processing through hyperbolic tangent and SoftMax functions to generate $A$. Subsequently, $A$ is multiplied with $V$ to produce an aggregated value tensor ($X'$).
\begin{align*}
A &= \text{SoftMax}(\tanh(QK^T)) \\
X' &= AV 
\end{align*}
The final output vector is the sum of the average-pooled $X$ and the $X'$ calculated within the attention block.
\begin{align*}
X'' &= \text{mean}(X)+ X'
\end{align*}
In MODEL-4, we employed two different attention blocks, represented as `ATTENTION LAYER 1' and `ATTENTION LAYER 2'. The former is applied on the embedding from the MPT model that takes the complete structure with the interpolated images (or the main-graph) as the input. The latter takes only a sub-structure corresponding to the initial, final, interpolated images, and their corresponding neighbors (or a sub-graph) as the input. For generating the sub-graph, we considered neighboring atoms of each image that were within a cut-off distance of 3~{\AA} from the corresponding image.  

\subsection{Model performance benchmarking}
\subsubsection{Scratch Models}
We refer to scratch models as those modified ALIGNN architectures which were not pre-trained on any bulk property prior to being trained for $E_m$ predictions. Thus, our scratch models provide a baseline for performance comparisons with our TL models. Specifically, we used two instances of the ALIGNN architecture to process two structures as input (i.e., the initial and final configurations, similar to MODEL-1), without any pre-training. The resultant embeddings from both instances were combined through addition (SCRATCH-ADD), subtraction (SCRATCH-SUB), or concatenation (SCRATCH-CC). We used this two-input-structure approach for training scratch models since the standard ALIGNN architecture is capable of taking only one input structure and any scratch resultant model will be incapable of identifying multiple migration pathways within the same structure.

\subsubsection{Classical ML models} 
To establish a performance benchmark in addition to scratch models, we compared our MPT model against classical ML models, namely a RFR and a GBR. Both classical models were constructed using a comprehensive feature set that originated from two sources: the matminer.featurizers package within the matminer library,\cite{ward2018matminer} and nine manually engineered features tailored to capture cation migration pathways. Matminer provided elemental, stoichiometric, electronic, environmental, structural, and interaction-based descriptors. Since RFR and GBR models cannot take two structures as input configurations simultaneously, we engineered the manual features to capture the characteristics of the migration pathway. The manual features included cation bond lengths, path distance, coordination number, and Voronoi polyhedra-derived attributes. The input structure to derive the features was the initial configuration of the migrating ion. From an initial set of 277 features, Pearson's correlation coefficient\cite{sedgwick2012pearson} was used for feature selection. While we removed features that were highly correlated with each other, only features exhibiting some correlation ($>$~0.10) with the target variable ($E_m$) were retained, resulting in a final set of 75 features. The hyperparameters of both RFR and GBR were optmized using five-fold cross-validation scores. Details on the specific features used and the optimized hyperparameters are compiled in Section~S2, Table~S2, and Table~S3 of the SI.  

\subsubsection{MACE-NEB E$_m$ calculation} 
In addition to scratch and classical models, we compare the performance of our TL models with $E_m$ predictions from NEB calculations, done with a universal MLIP, namely MACE-MP-0.\cite{batatia2025design,batatia2023foundation} For NEB calculations with MACE, we used the implementation available in the atomic simulation environment (ASE\cite{larsen2017atomic}). We performed the NEB calculations using two variations, one where we relaxed the endpoints using MACE-MP-0 before performing the NEB (we refer these predictions as MACE-NEB-RELAX or MACE-RELAX) and the other with no relaxation of the endpoints (MACE-NEB-NO-RELAX or MACE-NO-RELAX). Note that the MACE-NEB-NO-RELAX is the case that is similar to our TL model implementations, as we do not relax the endpoints before training the model, except for the charged state structures that were relaxed with DFT to obtain the true GS.\cite{devi2025Data} Specifics on the NEB calculations performed using MACE and ASE are described in Section~S3 and Table~S4 of the SI.

\section{Data availability}
All computed data and constructed models associated with this work are available online freely to all via our \href{https://github.com/sai-mat-group/predicting-migration-barriers}{GitHub} repository.

\section{Code availability}
All codes related to this work are available online freely to all via our \href{https://github.com/sai-mat-group/predicting-migration-barriers}{GitHub} repository. The source code of ALIGNN is available at the \href{https://github.com/usnistgov/alignn}{GitHub} repository maintained by the developers of ALIGNN.

\section{Acknowledgements}
G.S.G. and K.T.B. would like to acknowledge financial support from the Royal Society under grant number IES$\backslash$R3$\backslash$223036, and the United Kingdom Research and Innovation (UKRI) Engineering and Physical Sciences Research Council (EPSRC), under projects EP/Y000552/1 and EP/Y014405/1. G.S.G. acknowledges financial support from the Science and Engineering Research Board (SERB) of the Department of Science and Technology, Government of India, under sanction number IPA/2021/000007. R.D. thanks the Ministry of Human Resource Development, Government of India, for financial assistance. R.D. and G.S.G. acknowledge the computational resources provided by the Supercomputer Education and Research Centre, IISc, for enabling some of the calculations showcased in this work. We acknowledge National Supercomputing Mission (NSM) for providing computing resources of ‘Param Utkarsh’ at CDAC Knowledge Park, Bengaluru. PARAM Utkarsh is implemented by CDAC and supported by the Ministry of Electronics and Information Technology (MeitY) and Department of Science and Technology (DST), Government of India. Via our membership of the UK's HEC Materials Chemistry Consortium, which is funded by EPSRC (EP/X035859/1), this work used the ARCHER2 UK National Supercomputing Service (\href{http://www.archer2.ac.uk}{http://www.archer2.ac.uk}).

\section{Author Contributions}
G.S.G. and K.T.B. envisioned the project, supervised all aspects of the work, obtained funding and resources, and edited the manuscript. R.D. executed all aspects of the work, including data generation, data visualization, and writing the first draft of the manuscript.

\section{Competing Interests}
The Authors declare no competing financial or non-financial interests.


\bibliographystyle{unsrt}
\bibliography{sample}

\end{document}